  \def\selectedoptions{final}

\documentclass[
   \selectedoptions
  ]
  {aipproc}

\newcommand{\be}{\begin{eqnarray}}
\newcommand{\en}{\end{eqnarray}}
\newcommand{\no}{\nonumber}

\newcommand{\mc}{\mathcal}
\newcommand{\up}{u}
\newcommand{\dn}{d}
\newcommand{\upd}{u^{\dagger}}
\newcommand{\dnd}{d^{\dagger}}
\newcommand{\f}{f}
\newcommand{\p}{p}
\newcommand{\fd}{f^{\dagger}}
\newcommand{\pd}{p^{\dagger}}
\newcommand{\at}{\tilde{a}}
\newcommand{\atd}{\tilde{a}^{\dagger}}
\newcommand{\ad}{a^{\dagger}}

  \def\selectedlayoutstyle{6x9}

\layoutstyle\selectedlayoutstyle

\SetInternalRegister\hbadness{8000} 

\newcommand\doingARLO[2][]{%
  \ifx\mmref\undefined #1\else #2\fi
}
  
\begin{document}

\title 
      [A NEW SPINWAVE EXPANSION]
      {A new spinwave expansion for the ordered Kondo lattice}


\author{Nic Shannon}{
  address={Max--Planck--Institut f{\"u}r Physik komplexer Systeme,
N{\"o}thnitzer Str. 38, 01187 Dresden, Germany},
  email={shannon@mpipks-dresden.mpg.de},
  thanks={MPI--PKS etc.}
}

\copyrightyear  {2001}

\begin{abstract}
We present a concise introduction to a new spinwave 
expansion scheme for magnetically ordered Kondo lattice models.  
This is motivated by consideration of the ferromagnetically ordered 
phase of the ``double exchange'' system La$_{1-x}$Ca$_x$MnO$_3$.  
A brief overview is given of the consequences of quantum and thermal 
fluctuation effects for the magnetic properties of the double 
exchange ferromagnet.
\end{abstract}

\date{Vietri, 17th October 2001}

\maketitle

\section{Introduction}

The defining feature of transition metals is {\it mixed valence} --- 
their ability to donate a different numbers of electrons to chemical 
bonds according to the detailed charge balance of their environment.   
In addition, magnetic transition metals like Manganese have a strong 
Hund's first rule coupling.  This forces all electrons in the unfilled 
shell of the ion to 
align their spin, giving rise to a net local magnetic moment.

Taken together, these two facts have important consequences for 
magnetism in doped transition metal oxides, where the transition 
metal ions can frequently be found in states with different 
valence and therefore different total spin.  Where the electrons
are also able to move from site to site, local
moment behaviour normally associated with magnetic
insulators can coexist with metallic conduction.
In this article we will give a brief pedagogical overview
of a systematic method of describing the magnetic 
and charge excitations of such as a system.  
The approach developed has the conceptual and 
technical advantage that it does not introduce any artificial 
distinction between the magnetism of itinerant and localized electrons

\section{From material to model}

\begin{figure}[h]
  \includegraphics[height=0.25\textheight]{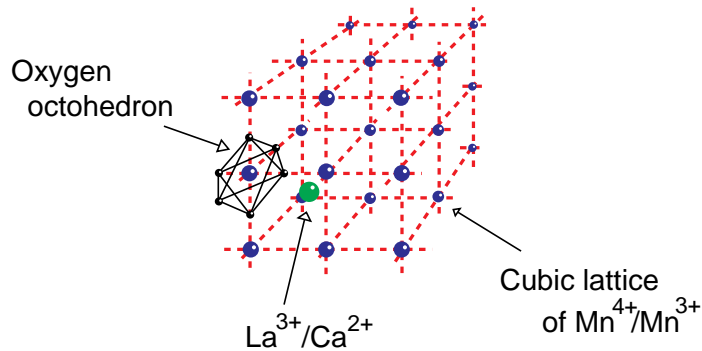}
\caption{Cubic lattice of Mn ions in La$_{1-x}$Ca$_x$MnO$_3$.}
\label{fig1}
\end{figure}

A good example of a mixed valent system with strong Hund's rule 
coupling is provided by the family of ``colossal magnetoresistance'' 
manganites La$_{1-x}$Ca$_x$MnO$_3$.  This compound has a perovskite
structure, like the High Temperature superconductors, and 
the magnetic manganese atoms occupy sites on a cubic lattice 
--- see Fig. \ref{fig1}.
We can rewrite the chemical formula of this manganite as 
(La$^{3+}$Mn$^{3+}$)$_{1-x}$(Ca$^{2+}$Mn$^{4+}$)$_x$(O$^{2-}$)$_3$.
Strong Hund's rule coupling then means that for each unit cell the 
system contains $1-x$ spin $S=2$ local moments (the Mn$^{3+}$ ions), 
and $x$ spin $S=3/2$ local moments (Mn$^{4+}$ ions).
States with an integer number of electrons per Mn site cannot conduct, 
and the parent compound LaMnO$_3$ is a spin $S=2$ antiferromagnetic (AF) 
insulator.

\begin{figure}[h]
  \includegraphics[height=0.25\textheight]{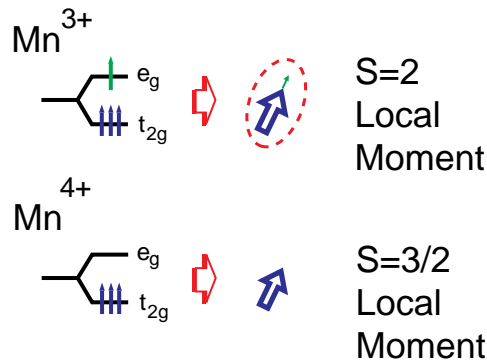}
\caption{Crystal field splitting and Mn spin states in 
La$_{1-x}$Ca$_x$MnO$_3$.}
\label{fig2}
\end{figure}

On doping with Ca, ``holes''
are introduced onto the Mn sites, so that the system now has
a mixture of $S=2$ Mn$^{3+}$ and $S=3/2$ Mn$^{4+}$ ions. 
The crystal field at the Mn sites splits Mn $d$--electron 
levels into a $t_{2g}$ and an $e_g$ multiplet, as shown in 
Fig. \ref{fig2}.  Where 
the $e_g$ electrons delocalize, this leads to 
a natural description of the material in terms of  
localized and itinerant electrons.  

The simplest model Hamiltonian we can use to describe such a system 
is the Kondo lattice model 
\be
\label{eqn:KondoH}
{\mc H}
  = -t \sum_{\langle ij \rangle\alpha} 
           c^{\dagger}_{i\alpha} c_{j\alpha}
       -J \sum_{i} \vec{S}_i.\vec{s}_i 
   &\qquad& 
\vec{s}_i = \frac{1}{2}\sum_{\alpha\beta} 
      c^{\dagger}_{i\alpha}
      \vec{\sigma_{\alpha\beta}}c_{i\beta}
\en
This model has already been much discussed during the course of this 
school, usually for small $S=1/2$ and in the limit
$
\mid J \mid \ll t
$
where $J < 0$ is the AF Kondo coupling.
Here we must consider the opposite limit of 
$ 
t \ll \mid J \mid
$
for ferromagnetic (FM) $J > 0$.  We will consider only
states in which the system is magnetically ordered, 
in which case it is reasonable to assume that the size
of the local moment $S \gg 1$.  For simplicity, we 
neglect the degeneracy of the itinerant electron $e_g$ orbital.

\begin{figure}[h]
  \includegraphics[height=0.25\textheight]{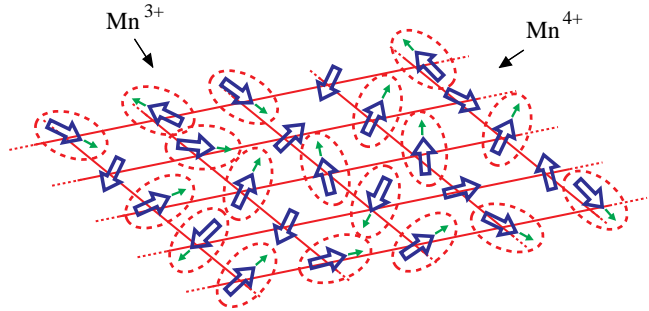}
\caption{Paramagnetic state in the absence of inter site hopping ($t=0$).}
\label{fig3}
\end{figure}

\begin{figure}[b]
  \includegraphics[height=0.2\textheight]{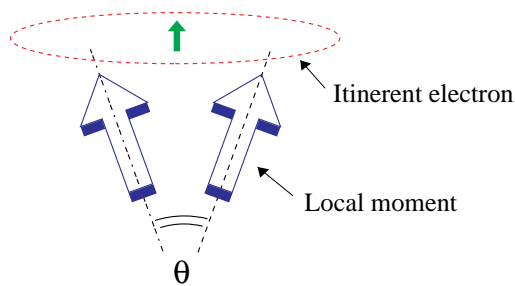}
\caption{A single double exchange bond in the limit 
$J\to \infty$, $S \to \infty$.}
\label{fig4}
\end{figure}

Since $t/J$ is a small parameter it makes sense to
diagonalize the Hund's rule coupling term $\propto J$ first, 
and then to reintroduce the kinetic energy $\propto t$.   
In the absence of any 
coupling between sites ($t=0$), for any density of holes, 
the system must be an insulator with paramagnetic Curie law 
susceptibility --- see Fig \ref{fig3}.  
For finite $t > 0$ the ground state, at least at large $S$, will be a 
metallic FM, since the band kinetic energy of the itinerant electrons
is greatest when all the spins are aligned --- see Fig \ref{fig4}.
This effect is was named ``double exchange'' by Zener \cite{zener}, 
because in La$_{1-x}$Ca$_x$MnO$_3$, itinerant electrons move between 
from one Mn site to another through connecting 0$_{2p}$ orbitals. 
Where such a FM groundstate is realized, it is therefore referred to as
a {\it double exchange ferromagnet} (DEFM).  

In what follows we will develop a controlled expansion scheme for 
calculating the properties of the localized/itinerant electron 
system described by Eqn. \ref{eqn:KondoH}.  
Here we will only discuss the DEFM, but the methods developed 
are much more general and may be applied to any magnetically ordered 
Kondo lattice model.

\section{Semi--classical treatment}

\begin{figure}[h]
  \includegraphics[height=0.75\textheight]{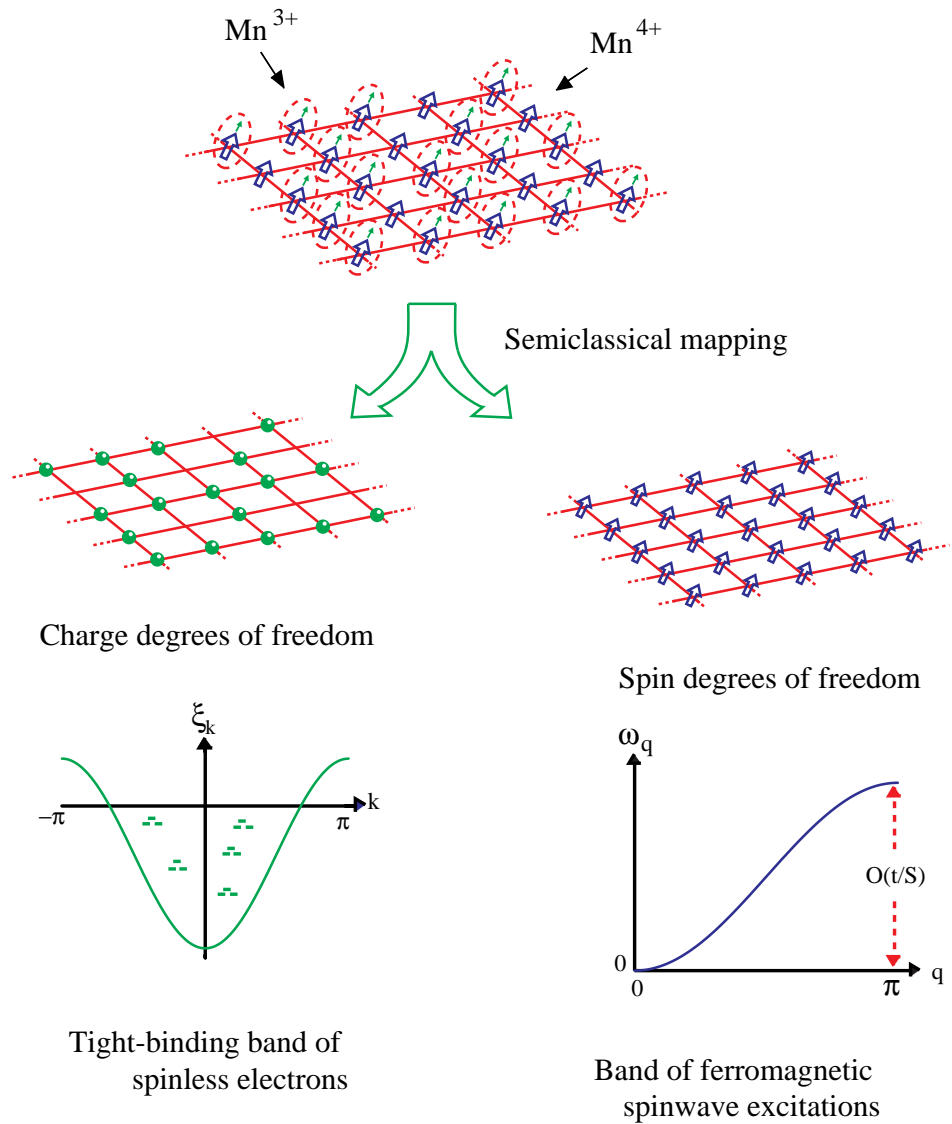}
\caption{DEFM for $t>0$ and semiclassical mapping
onto independent spin and charge excitations in limit 
$J\to \infty$, $S \to \infty$.}
\label{fig5}
\end{figure}

The physics of the DEFM can be understood easily at a 
semi--classical level, by analogy with the problem of two local
moments sharing a single electron.  This was first solved
by Anderson and Hasegawa \cite{anderson}, who considered
the Hamiltonian Eqn. \ref{eqn:KondoH} restricted to
two sites.  In the limit $S \to \infty$ we can
define an angle $\theta$ between the two local
moments (Fig. \ref{fig5}).  The energy of the
ground state for a given $\theta$ is given by  
\be
E_0 &=& -\frac{JS}{2} - t \cos\left(\frac{\theta}{2}\right)
\en
where an $SU(2)$ rotation of the local spin quantization axis, 
together with the limit $J \to \infty$, has been used 
to project out those electron states which are not aligned with
the local moment on each site.
There are two contributions to this an energy, a term proportional 
to  $J$ which is the Hund's first rule binding energy of an electron 
on a single site, and a term proportional to $t$ which is the kinetic
energy gain which comes from sharing the electron between two sites.
The factor of $\cos(\theta/2)$ comes from the $SU(2)$ rotation. 
Using the relation $\cos\theta = \vec{S}_1.\vec{S}_2/S^2$, 
and a standard trigonometric identity for double angles
we can write the energy of this state as
\be 
E_0 &\approx& - \frac{t}{4S^2} \vec{S}_1.\vec{S}_2
  + const. 
\en
The kinetic energy of the itinerant electron generates
an effective FM Heisenberg exchange interaction 
$
J_{DE} = t/4S^2
$
between the two sites.
The generalization of this semi--classical analysis to the lattice
\cite{degennes}, gives an effective Hamiltonian for the DEFM 
\be
\label{eqn:effectiveH}
{\mc H}_{eff} = {\mc H}_{\rho} + {\mc H}_{\sigma}
\en
with independent spin and charge excitations
\be
{\mc H}_{\rho} = -t \sum_{\langle ij \rangle}\fd_i\f_j 
   \qquad
{\mc H}_{\sigma} = -J_{DE} \sum_{\langle ij \rangle}
   \vec{S}_i.\vec{S}_j
\en
Here the sum $\sum_{\langle ij \rangle}$ runs over nearest neighbour 
sites, and $\fd_i$ is the creation operator for an electron 
aligned with the local moment on site $i$.
The charge part of the Hamiltonian describes electrons 
with a tight binding band dispersion 
$
\epsilon_k = -zt \gamma_k 
$
and the spin Hamiltonian has spinwave
excitations with dispersion
$
\omega_q = zJ_{DE}S [ 1 - \gamma_q]
$ 
where 
$
\gamma_k = (\cos k_x + \cos k_y + \cos k_z)/3
$
is the structure factor for the cubic lattice of Mn sites
in units where the lattice constant $a=1$.
The size of the effective exchange interaction 
$J_{DE}$ is determined by the ground state average of the 
itinerant electron electron kinetic energy
in the FM ground state of the model, and so is also 
proportional to $t$  
\be
J_{DE} &=& \frac{t}{4S^2} \sum_{\langle ij \rangle}
   \langle \fd_i\f_j \rangle_0
\en
In fact the energy scales of both spin and charge excitations {\it must} 
be be proportional to $t$, since the role of hopping is to lift the 
degeneracy between the many different charge configurations for a 
given doping and spin state.

Using this effective Hamiltonian, we can understand both the 
metallic conduction of the FM phase of the CMR Manganites, 
and the fact that their spinwave spectrum, when measured
in inelastic neutron scattering experiments,
looks a lot like that of a FM nearest neighbour 
Heisenberg model.  However we have neglected all quantum 
fluctuation effects, which means we cannot explain, 
for example,  the large damping of spinwaves which is also 
observed in Neutron scattering experiments.  The most important 
feature missing from this Hamiltonian is the
{\it interaction} between spin and charge excitations,
which is present when fluctuation effects are
taken into account.   It is this interaction vertex 
which we now calculate.   However, in order to do this,
we must first re--derive the semi--classical treatment
of the DEFM in a rigorously controlled manner.
Since the literature on the DE model is too large to review
in this format,
we will not make any attempt to discuss the alternative
approaches which might be taken to this problem.

\section{New spinwave expansion}

Since the Hund's rule coupling $J$ is much larger than the 
kinetic energy scale $t$, we begin by setting $t=0$ 
and considering the atomic limit of the Kondo lattice Hamiltonian 
Eqn.~\ref{eqn:KondoH}. 
It is our aim to first diagonalize the local Kondo interaction, 
before reintroducing the kinetic energy terms. 
Physically, it makes no sense to distinguish between the 
magnetic excitations of the local moment $\vec{S}$ and
the spin of the itinerant electron $\vec{s}$, so
we consider instead the total spin $\vec{T} = \vec{S} + \vec{s}$.
In terms of this coordinate, the Hamiltonian Eqn. \ref{eqn:KondoH}
for a single sight can be written
\be 
-J \vec{S}.\vec{s} 
   = -\frac{J}{2}\left[\vec{T}^2 - \vec{S}^2 - \vec{s}^2\right]
   = -\frac{J}{2}\left[T(T + 1) - S(S+1) - s(s+1)\right]
\en
Here $S$ is the size of the local moment, $s$ is the spin quantum 
number for the itinerant electron, and the total spin quantum number 
$T$ can take on values $T = S \pm s $.  There are four possible
states.
The lowest energy state has the eigenvalue $E = -JS/2$.  It corresponds to 
one itinerant electron being perfectly aligned with the local moment 
to give a total spin $T = S+1/2$. 
There are two states with eigenvalue $E=0$ and total spin $T = S$.
In these the itinerant electron orbital is either empty or doubly
occupied.
There is also a high energy state with eigenvalue $E = J(S+1)/2$,
and total spin eigenvalue $T = S-1/2$.
This  has one itinerant electron perfectly ``anti--aligned'' with the 
local moment.

\begin{figure}[h]
  \includegraphics[height=0.3\textheight]{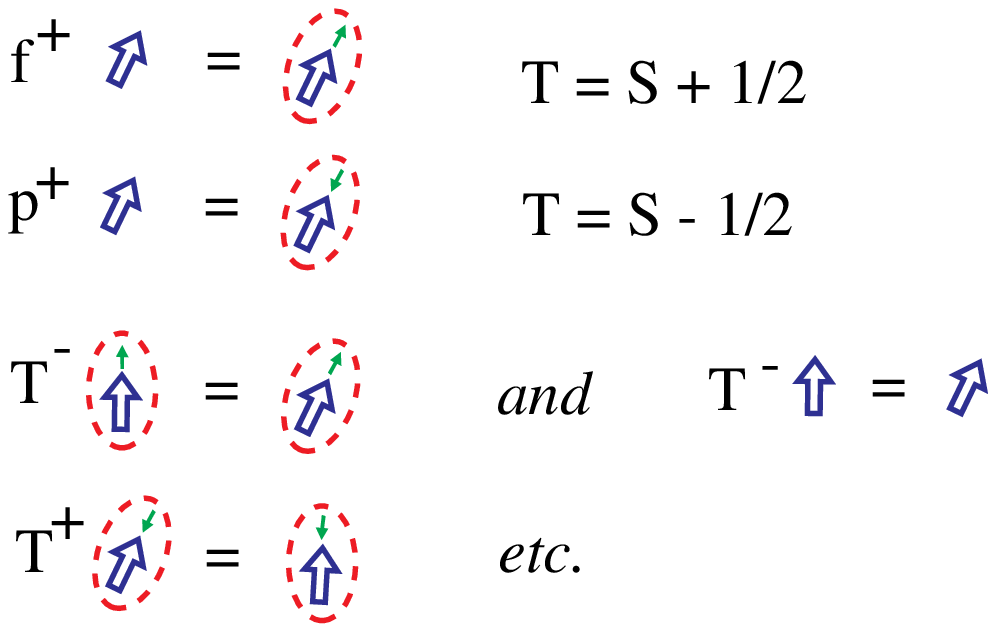}
\caption{Cartoon illustration of new operator coordinates.}
\label{fig6}
\end{figure}

\begin{figure}[h]
  \includegraphics[height=0.25\textheight]{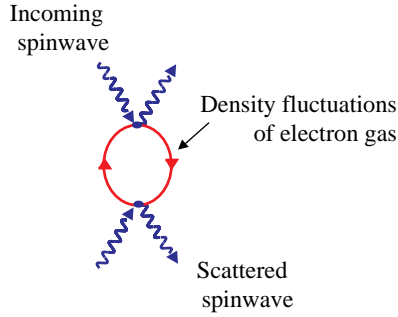}
\caption{Interaction between spinwaves mediated by density
fluctuations of the electron gas.}
\label{fig7}
\end{figure}

Now we introduce electron operators which create
these states, when acting on a vacuum in which the local
moment has some definite $z$--component $S^z$.  We introduce
one Fermi operator $\{\f,\fd\}=1$ to describe high spin
states, and another Fermi operator $\{\p,\pd\}=1$ to 
describe low spin states.  These are illustrated in 
cartoon form in Fig. \ref{fig6}.
Next we need to introduce a bosonic description of the spin 
algebra of the total spin $\vec{T}$.  We do this by 
generalizing the usual Holstein--Primakoff 
transformation used to describe rotations to spins of a fixed length 
\cite{holstein} to the case in which the length of the spin is an operator.
\be
T^z = &S^z + \frac{\upd \up - \dnd \dn}{2}&
  = T - \atd\at\\
T^+ = &S^+ + \upd\dn&
  = \sqrt{2T} 
   \sqrt{1 - \frac{\atd\at}{2T}} \at
\en
where
\be
T = S + \frac{\fd\f -\pd\p}{2}
\en
The physical meaning of these spin operator is 
illustrated in Fig. \ref{fig6}.
Finally, we expand the operators $\{\f,\p,\at\}$ in terms 
of the operators $\{\up,\dn,a\}$, where $[a,\ad] =1$ is 
the Holstein--Primakoff Boson for the core spin $\vec{S}$.
We do this order by order in the small parameter $1/\sqrt{S}$.
We must make sure that all the correct canonical
(anti--)commutation relations are conserved by this 
transformation {\it at each order}.  
At the end of all this hard work, we are left with 
an {\it inverse} transformation 
\be
\label{inverse}
c_{\uparrow} &=& \f\left(1 - \frac{\atd\at + \pd \p}{4S}\right)
- \frac{\p \atd}{\sqrt{2S}} + {\mc O}(1/S^{3/2}) \no \\
c_{\downarrow} &=& \p\left(1 - \frac{\atd\at + \f \fd}{4S} \right)
+ \frac{\f \at}{\sqrt{2S}} + {\mc O}(1/S^{3/2})
\en
which can be substituted into the kinetic energy term
in Eqn. \ref{eqn:KondoH} to give the interaction between
spin and charge excitations.  
In the limit $J \to \infty$ the high energy ``$\p$''
states can be neglected entirely.  In terms of the 
new coordinates, the effective Hamiltonian 
Eqn. \ref{eqn:effectiveH} reads
\be
{\mc H}_{\rho} = \sum_{k}\epsilon_k \fd_k\f_k 
   \qquad
{\mc H}_{\sigma} = \sum_{q}\omega_q \atd_q \at_q
\en
The spin and charge modes now interact according to
\be
\label{interaction}
{\mc V} &=& 
   \frac{1}{N} \sum_{k_1\ldots k_4}  v^{13}_{24}
   : \fd_1 \f_2 : \atd_3 \at_4 \delta_{1+3-2-4} + \ldots\\
&v^{13}_{24}& =  \frac{1}{4(S+\frac{1}{2})}
   \left[ 
   \left( 1 + \frac{1}{8S} \right) 
   \left( \epsilon_{1+3} + \epsilon_{2+4}\right)
   - \left(\epsilon_1 + \epsilon_2\right)
   \right]
\en
Here the charge density $: \fd \f :$ is measured
relative to the average density of itinerant electrons,
and corrections to the vertex at ${\mc O}(1/S^2)$ have
been suppressed.
The form of the vertex is physically correct --- 
spin and charge quantum numbers are independently conserved
(as they must be in the limit $J \to \infty$), and 
the strength of their interaction is proportional to $t$.
The consequence of this vertex is an effective interaction
between charge and spin modes at ${\mc O}(1/S^2)$, shown 
in Fig \ref{fig7}.
These results were first introduced in \cite{EPL}.
Further details of the
transformation, including terms up ${\mc O}(1/S^2)$, 
are given in \cite{shannon} and \cite{PRB}.

\section{Results for the Double Exchange Ferromagnet}

We now briefly illustrate some of the consequences of 
the interaction between spin and charge modes for the 
physical properties of the DEFM.  All examples
given are for a system with a density of $x=0.3$ holes
per site, spin $S=3/2$ and hopping integral $t=1$.
These values are chosen to correspond with 
the FM state of La$_{1-x}$Ca$_x$MnO$_3$ with 
maximumal $T_c$.  Since the Hamiltonian is 
quadratic in Fermi and Bose fields, and the commutation
relatations of all operators are well defined, standard 
diagramatic techniques can be used to calculate the spin 
and charge response functions of the model.

\begin{figure}[h]
  \includegraphics[height=0.25\textheight]{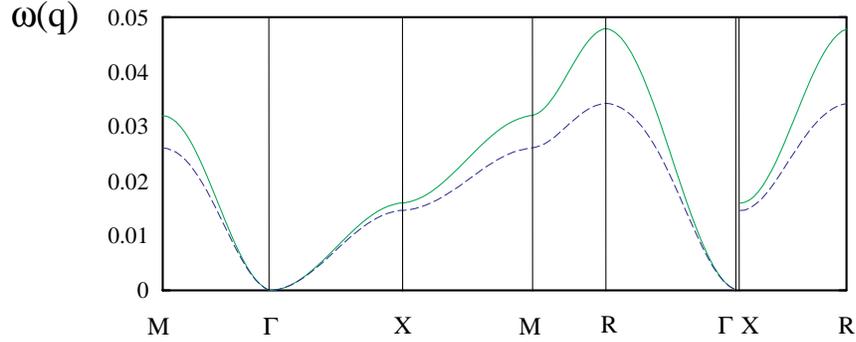}
\caption{Spinwave spectrum of the DEFM at zero temperature.  
Upper solid line --- semi--classical result throughout the Brillouin
zone.  Lower dashed line --- 
result when leading quantum corrections are taken 
into account.}
\label{fig8}
\end{figure}

In Fig.~\ref{fig8} the semi--classical spinwave
spectrum $\omega_q$ of the DEFM is shown, together with the
renormalized spectrum when the leading quantum
corrections to the spinwave dispersion at are taken
into account.  The effect of the interaction is
large, especially at the zone boundary.  The spinwave
over all scale of spinwave dispersion is reduced, and 
no longer has the simple form associated
with a nearest neighbour Heisenberg FM.  
Physically, this is because quantum 
fluctuations about the FM groundstate generate 
effective non nearest neighbour exchange couplings.

\begin{figure}[h]
  \includegraphics[height=0.25\textheight]{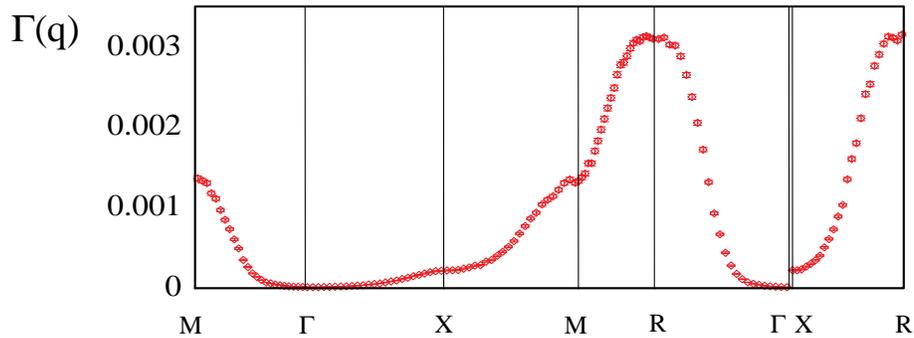}
\caption{Damping of spinwaves in the DEFM throughout the 
Brillouin zone.}
\label{fig9}
\end{figure}

Spinwaves excitations are exact eigenstates of the Heisenberg
FM, but are damped in the DEFM.  This damping arises because
spinwaves can decay into lower energy spin excitations, dressed
with charge density fluctuations (electron--hole pairs).
This damping is also large, as is illustrated in Fig.~\ref{fig9}

\begin{figure}[h]
  \includegraphics[height=0.35\textheight]{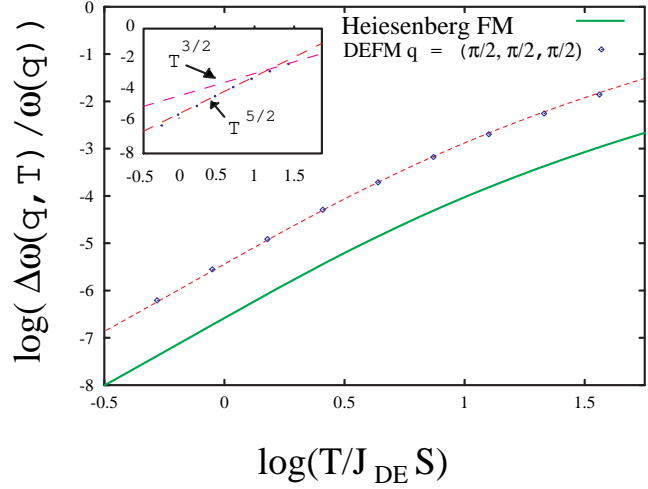}
\caption{Log--log plot of finite temperature corrections to 
the spinwave spectrum of the DEFM, together with those
of the semi--classically equivalent Heisenberg FM.  
Temperature corrections
$\Delta\omega(q,T)$ are normalized to bare dispersion $\omega(q)$,
temperatures to the scale of spinwave dispersion $J_{DE}S$.
}
\label{fig10}
\end{figure}

Temperature corrections to the spinwave dispersion in 
a Heisenberg FM are exactly proportional to the zero 
temperature dispersion, and scale as $T^{5/2}$ at
low temperatures.  Temperature corrections in the DEFM
are substantially enhanced, compared with the semi--classically
equivalent Heisenberg FM, and are nolonger simply 
proportional to the zero temperature dispersion.
The comparison of the two models is illustrated in Fig.~\ref{fig10}.
Detailed results will be presented in a forthcoming publication 
\cite{PRB}.

\section{Conclusions}

We have developed a controlled large $S$ expansion of the 
Kondo Lattice model, based on local eigenstates of total
spin.  This approach can applied to any magnetically 
ordered state, but is particularly convenient when
working in the ``double exchange'' limit $J \gg t$.
This approach has been applied to the double exchange ferromagnet,
which has been shown to behave in a quite different way 
from the Heisenberg ferromagnet when quantum effects are 
correctly included.

\begin{theacknowledgments}
It is our pleasure to acknowledge the collaboration 
of Andrey V.\ Chubukov in this work, and support under 
NSF grant DMR--9632527 and the 
visitor's program of MPI--PKS Dresden.
We would like also to express our gratitude to 
Adolfo Avella and the organizers of the 
Vietri training course.
\end{theacknowledgments}


\doingARLO[\bibliographystyle{aipproc}]
          {\ifthenelse{\equal{\AIPcitestyleselect}{num}}
             {\bibliographystyle{arlonum}}
             {\bibliographystyle{arlobib}}
          }
\bibliography{salerno}

\end{document}